\documentstyle[preprint,aps,tighten]{revtex}

\begin{document}
\draft
\title{Astrophysical factors:Zero energy vs. Most effective energy}

\author{Theodore E. Liolios $^{1,2,3}$ \footnote{theoliol@physics.auth.gr}}

\address{$^1$Institute of Physics and Astronomy, University of Aarhus, DK-8000 Aarhus C, Denmark\\
\footnote{Correspondence address}$^2$University of Thessaloniki, Department of Theoretical Physics,
Thessaloniki 54006, Greece\\
$^3$ Hellenic War College, BST 903, Greece\\}
\maketitle
\begin{abstract}
Effective astrophysical factors for non-resonant astrophysical nuclear
reaction are usually calculated with respect to a zero energy limit. In
the present work that limit is shown to be very disadvantageous compared to
the more natural effective energy limit. The latter is used in order to
modify the thermonuclear reaction rate formula in stellar evolution codes 
so that it takes into account both plasma and laboratory screening effects.
\end{abstract}


\oddsidemargin -0.25cm \evensidemargin -0.25cm \topmargin -1.0cm \textwidth %
16.3cm \textheight 22.3cm

Nuclear astrophysics experiments are characterized by their continuous
effort towards the measurement of the zero energy astrophysical factor $%
\left( AF\right) \,$ which is used in the thermonuclear reaction rate
formulas. Recent and past experiments\cite{kraussdd,lunasecond} have all
attempted to lower the energy of the experiment in order to avoid the
inaccuracies incurred by the inevitable extrapolation to zero energies. That
action, however, gives rise to an undesirable electron screening effect
which enhances low energy data (\cite{lioliosepj} and references therein).
Moreover, experimentalists often fit their $AF$ data with a zero-energy
Taylor polynomial which will be shown to be less suitable than the more
natural effective-energy one. The results of this paper should be viewed
as an improvement (for some cases)in the theoretical extraction of the astrophysical factor and 
not as a correction of the way experimentalists present their results.
In the formulas below we follow the formalism
and notation of Ref. \cite{claytonbook}.

The effective astrophysical factor $\left( EAF\right) $ $S_{eff}$ which
appears in the non-resonant thermonuclear reaction rate 
\begin{equation}
r_{12}=\frac{7.20\times 10^{-19}}{1+\delta _{12}}\frac{N_{1}N_{2}}{%
AZ_{1}Z_{2}}f_{S}\tau ^{2}e^{-\tau }S_{eff}  \label{r12}
\end{equation}
is defined as

\begin{equation}
S_{eff}=\sqrt{\frac{\tau }{4\pi }}\frac{e^{\tau }}{E_{0}}\int_{0}^{\infty
}S\left( E\right) \exp \left[ -\left( \frac{E}{kT}+2\pi n\right) \right]dE
\end{equation}
and can be calculated by expanding the astrophysical factor $S\left(
E\right) $ either around zero or around the most effective energy of
interaction $E_{0}.$ In the first case the expansion

\begin{equation}
S\left( E\right) =\sum_{n=0}^{\infty }\frac{1}{n!}S^{\left( n\right) }\left(
0\right) E^{n}  \label{szero}
\end{equation}
yields(\cite{bahcallbook},\cite{ueda})

\begin{equation}
S_{eff}^{\left( 0\right) }=\sum_{n=0}^{n_{M}}\frac{1}{n!}S^{\left( n\right)
}\left( 0\right) E_{0}^{n}\sum_{k=0}^{k_{M}}\frac{P_{2k}\left( n\right) }{%
k!\left( 12\right) ^{k}\tau ^{k}}  \label{seffzero}
\end{equation}
while an expansion around the Gamow peak energy $E_{0}$

\begin{equation}
S\left( E\right) =\sum_{n=0}^{\infty }\frac{1}{n!}S^{\left( n\right) }\left(
E_{0}\right) \left( E-E_{0}\right) ^{n}  \label{sgamow}
\end{equation}
yields(\cite{bahcallbook},\cite{ueda})

\begin{equation}
S_{eff}^{\left( G\right) }=\sum_{n=0}^{n_{M}}\frac{1}{n!}S^{\left( n\right)
}\left( E_{0}\right) E_{0}^{n}\sum_{r=0}^{n}\left( -1\right) ^{r} \left(%
\frac{n}{r}\right) \sum_{k=0}^{k_{M}}\frac{P_{2k}\left( n-r\right) }{%
k!\left( 12\right) ^{k}\tau ^{k}}  \label{seffgamow}
\end{equation}
where the polynomials are given in Ref. \cite{ueda}. For example, the first
two are 
\begin{equation}
P_{0}\left( n\right) =1,\,P_{2}\left( n\right) =12n^{2}+18n+5
\end{equation}
In most nuclear astrophysics experiments of non-resonant reaction
experimentalist\cite{rolfsbook} use a second order truncation of Eq. $\left( 
\ref{szero}\right) $ as a fitting formula. They obtain the values $S\left(
0\right) ,S^{\left( 1\right) }\left( 0\right) ,S^{\left( 2\right) }\left(
0\right) $ which are then inserted into a second order truncation of Eq. $%
\left( \ref{seffzero}\right) .$

However, zero energy proximity has serious disadvantages since at such very
low energies, which are extremely difficult to attain, the experiment is
hampered by such effects as beam instabilities, impurities, electron
screening effects, very small cross sections etc. On the other hand the
Gamow peak polynomial of Eq. $\left( \ref{sgamow}\right) $ is hardly ever
used although it leads to a more accurate calculation of the $EAF$. In Ref. 
\cite{ueda} it was noted that the accuracy obtained by the first term of Eq. 
$\left( \ref{seffgamow}\right) $, that is $S_{eff}^{\left( G\right) }\simeq
S\left( E_{0}\right) $ , is equivalent to the accuracy achieved with
knowledge of $S\left( 0\right) $ and its derivatives. That, of course, is
also obvious from Eq. $\left( \ref{szero}\right) \,$which gives

\begin{equation}
S\left( E_{0}\right) =\sum_{n=0}^{\infty }\frac{1}{n!}S^{\left( n\right)
}\left( 0\right) E_{0}^{n}
\end{equation}
We can further elaborate on that competition of Eq. $\left( \ref{seffgamow}%
\right) $ versus Eq. $\left( \ref{seffzero}\right) $ by disregarding the
associated derivatives of the AFs, an approximation which doesn't cause any
significant error\cite{jennings}. In that case we can write

\begin{equation}
S_{eff}^{\left( G\right) }=S_{eff}^{\left( 0\right) }+\left( 1+\frac{5}{%
12\tau }\right) \sum_{n=1}^{\infty }\frac{1}{n!}S^{\left( n\right) }\left(
0\right) E_{0}^{n}
\end{equation}
where it is now obvious that $S_{eff}^{\left( G\right) }$ incorporates $%
S_{eff}^{\left( 0\right) }$ along with an infinite number of corrective
terms. It is again obvious that a simple first order correction to the
constant Gamow peak AF $S\left( E_{0}\right) $ is equivalent to knowing an
infinite number of zero energy derivatives.

On the other hand the crucial region of energies is not close to zero but
the Gamow window $\left[ -\Delta /2+E_{0},E_{0}+\Delta /2\right] ,\,$with $%
\Delta $ being as usual the full width at $1/e.$ Consequently, while in the
case of $S_{eff}^{\left( 0\right) }$ we attempt to lower the energy as close
to zero as possible, paying a great price, when $S_{eff}^{\left( G\right) }$
is used the situation improves considerably. First, we only have to perform
measurements inside the Gamow window, thus avoiding the dangerous zero
energy region. Then we obtain accurate results by including a limited number
of terms while in the zero-energy formulas a second order expansion is not
adequate as measurements are far away from the origin. It is surprising that some 
experimentalist have been satisfied with fitting a second order expansion of
the AF given by Eq. $\left( \ref{szero}\right) $ to their data. It is easy
to show that including some more terms will result in different zero energy
AFs. Fortunately the results that are used in stellar
calculations are often based on independent analyses of the
experiments (often with theoretical guidance rather than the
naive Taylor expansion)

Another positive aspect of adopting Eq. $\left( \ref{sgamow}\right) $
instead of Eq. $\left( \ref{szero}\right) $ is the electron screening
effect, which enhances all very low energy data. The lower the energy, the
more pronounced the effect which makes Eq. $\left( \ref{sgamow}\right) $
seem all the more attractive. In fact, in the above formulas, the quantities 
$S^{\left( n\right) }\left( 0\right) $ and $S^{\left( n\right) }\left(
E_{0}\right) $ are referring to bare nuclei. However, in the laboratory $%
\left( L\right) $ the experimental $\left( ex\right) $ values $%
S_{ex}^{\left( n\right) }\left( 0\right) $ and $S_{ex}^{\left( n\right)
}\left( E_{0}\right) $ are enhanced as they are multiplied by the screening
enhancement factor $\left( SEF\right) $ given by\cite{shoppaatomic}

\begin{equation}
f_{L}\left( E\right) =\exp \left( \pi n\frac{U_{e}^{L}}{E}\right)
\label{sef}
\end{equation}
where $n$ is the Sommerfeld parameter, and $U_{e}^{L}$ is the screening
energy obtained via theoretical models. For nuclei involved in the CNO
bi-cycle or more advanced burning stages $U_{e}^{L}$ can be given by\cite
{lioliosluna}

\begin{equation}
U_{TF}^{SL}=-32.9Z_{1}^{4/3}Z_{2}\,eV  \label{utfe}
\end{equation}
where $Z_{1},Z_{2}$ are the atomic numbers of the target and the projectile
respectively.
Alternatively, if one wants to fully explore the thermal, ionization and relativistic effects 
on screening more elaborate formulas can be used\cite{lioliossef}. 
The bare-nucleus AF $S\left( E\right) $ is related to the corresponding
experimental value $S_{ex}\left( E\right) $ by the formula:

\begin{equation}
S\left( E\right) =S_{ex}\left( E\right) f_{L}^{-1}\left( E\right)
\end{equation}
Another disadvantage of the zero-energy AF is that we cannot obtain $S\left(
0\right) $ from the above formula, whereas the Gamow-peak AF is readily
given by $S\left( E_{0}\right) =S_{ex}\left( E_{0}\right) f_{L}^{-1}\left(
E_{0}\right) $. Disregarding second order derivatives $S^{\left( 2\right)
}\left( E_{0}\right) $, which play a minor role and their correction (or
their inclusion according to Ref. \cite{jennings}) would be pointless we have

\begin{equation}
S_{eff}^{\left( G\right) }\simeq f_{L}\left( E_{0}\right) S_{ex}\left(
E_{0}\right) \left[ 1+\frac{5}{12\tau }+\frac{5}{2\tau }\frac{S_{ex}^{\left(
1\right) }\left( E_{0}\right) }{S_{ex}\left( E_{0}\right) }E_{0}+\frac{15}{%
12\tau }\ln f_{L}\left( E_{0}\right) \right]  \label{sls}
\end{equation}
Actually, neglecting the screening effect is equivalent to setting $%
f_{L}\left( E\right) =1.$

The above formula vividly depicts that the value of the $EAF$ used in Eq. $%
\left( \ref{r12}\right) $ actually depends on the $SEF$ and the model used
to describe the screening effect. This of course is known empirically to
experimentalists who try to correct their data in the laboratory. What is
derived here for the first time is the analytic dependence of the EAF on the
screening models selected each time. And there are many, indeed. Another
novelty of the above formula is that it eliminates all the tedious
corrections that are necessary during nuclear astrophysics experiments.
Moreover, it corrects all nuclear astrophysics experiments performed in the
past, which have disregarded the screening effect.

There is another major advantage of Eq. $\left( \ref{sls}\right) $ that
needs to be underlined. By now it is obvious that, if we disregard screening
effects, the $EAF$ is approximately equal to either $S\left( E_{0}\right) $
or $S\left( 0\right) .$ Therefore, if the latter is selected by means of Eq. 
$\left( \ref{seffzero}\right) $, all temperature effects of the $EAF\,$are
actually neglected, since $S\left( 0\right) \,$will be the same along the
whole profile of the star. On the contrary, if the former is selected then
the $EAF$ takes into account the region of the star where the reaction takes
place via the temperature dependence of the most effective energy of
interaction $E_{0}\left( T\right) .$ For example, if we adopt Eq. $\left( 
\ref{seffzero}\right) $ in our solar codes using the bare-nucleus value\cite
{lunasecond} $S\left( 0\right) =5.32$ for the $He^{3}\left( He^{3},2p\right)
He^{4}$ reaction then that will practically be the value of the $EAF$ at the
center of the sun, throughout its energy production core as well as in any
other stellar interior no matter how hot it is. That is indeed an unnatural
result. The only way Eq. $\left( \ref{seffzero}\right) \,$could
counterbalance the temperature dependence of the leading term $S\left(
E_{0}\right) $ of Eq. $\left( \ref{seffgamow}\right) $ is to include an
infinite number of temperature dependent terms, which is of course futile
since that can be accomplished by merely setting $S_{eff}\simeq S\left(
E_{0}\right) \,$in the reaction rate.

To gain an idea of the screening corrections to the $EAF$ let us consider
the break-up reaction $He^{3}\left( He^{3},2p\right) He^{4}$ whose $SEF$
according to a recent model\cite{lioliosluna} is $f_{L}\left( E_{0}\right)
=1.23.$ The screening correction inside the brackets in Eq. $\left( \ref{sls}%
\right) $ is roughly $\left( 15/12\tau \right) \ln f_{L}\left( E_{0}\right)
=4\times 10^{-3}$ while the first significant term is $\left( 5/12\tau
\right) =7.7\times 10^{-3}.$ Obviously, the screening correction is
comparable to the first corrective term which is usually retained in the
formula of the $EAF$. Of course they are both negligible and the only
significant correction in Eq. $\left( \ref{sls}\right) $ is that of the SEF
multiplied by $S_{ex}\left( E_{0}\right) .$ As for reactions involved in
more advanced burning stages than the $pp$ one we can consider the most
important reaction of the $CNO$ bi-cycle namely $N^{14}\left( p,\gamma
\right) O^{15}$. For that reaction the two previously mentioned corrections
are roughly the same $\left( \sim 6.\,\times 10^{-3}\right) $. Note that
including any of the corrective terms inside the brackets would be
meaningless for another reason. As we observed their contribution is of
order $1\%$ while the experimental error of the leading terms $S\left(
E_{0}\right) $ (or even $S\left( 0\right) $ when Eq. $\left( \ref{seffzero}%
\right) $ is adopted) is much larger\cite{bahcallbook}.

We can now safely argue that all screening effects on the EAF can be taken
into account by multiplying the (uncorrected) $EAF$ with the $SEF$ given by $%
f_{L}\left( E_{0}\right) .$ Therefore the reaction rate itself is now
multiplied by a $SEF$ which is the combination of laboratory and plasma
screening effects. The reaction rate is now written

\begin{equation}
r_{12}=\frac{7.20\times 10^{-19}}{1+\delta _{12}}\frac{N_{1}N_{2}}{%
AZ_{1}Z_{2}}f_{SL}\tau ^{2}e^{-\tau }S_{eff}^{\left( G\right) }
\label{r12sl}
\end{equation}
where $S_{eff}^{\left( G\right) }\,$is merely a single experimental
measurement, that is $S_{eff}^{\left( G\right) }\simeq S_{ex}\left(
E_{0}\right) ,$ and the combined $SEF$ $f_{SL}$ is given by

\begin{equation}
f_{SL}\left( E_{0}\right) =\exp \left[ \frac{\pi n\left( E_{0}\right) }{E_{0}%
}\left( U_{e}^{S}-U_{e}^{L}\right) \right]   \label{fsl}
\end{equation}
Note that the values of the screening energies $\left(
U_{e}^{S},U_{e}^{L}\right) $ in the above formula are absolute ones. At
first sight it seems that Eq. $\left( \ref{r12sl}\right) \,$and Eq. $\left( 
\ref{fsl}\right) \,$indicate that $r_{12}\neq r_{21\text{\thinspace }}$, 
since $U_{e}^{L}\left( Z_{1},Z_{2}\right) \neq U_{e}^{L}\left(
Z_{2},Z_{1}\right) $ , an error also committed in the plasma $SEF$ derived
in Ref. \cite{cameron}. This is not the case here. The combined $SEF\,$given
by $f_{SL}$ must always be  coupled with the appropriate experimental value $%
S_{ex}\left( E_{0}\right) \,$so that it is always $r_{12}=r_{21\text{%
\thinspace }}.$

According to Eq. $\left( \ref{fsl}\right) $, plasma and laboratory screening
have opposite effects on the reaction rate. The theoretical value of $%
U_{e}^{L}$ has already been discussed. On the other hand the value of the
plasma screening energy can be derived using various theoretical models such
as Salpeter's\cite{salpeter} according to which $
U_{e}^{S}=Z_{1}Z_{2}e^{2}r_{D}^{-1}$ where $r_{D}$ the Debye radius, or even
Shaviv's\cite{shaviv} according to which $U_{e}^{S}$ =$%
1.5Z_{1}Z_{2}e^{2}r_{D}^{-1}.$

Regarding these two plasma screening prescriptions a comment is imperative.
In a recent work\cite{bahshav}, Shaviv's prescriptions has been argued
against. However, the relation between Salpeter's approach and Shaviv's is
as it should be. Actually they provide upper and lower limits for the
screening energies, as has often been done in laboratory screening models%
\cite{lioliosluna} where a sudden (Salpeter's) and an adiabatic (Shaviv's)
limit is defined. It is important to realize that the most accurate way of
deriving a $SEF$ is to provide a sudden and an adiabatic limit with a
minimum discrepancy between them. The two formulas in question fulfill that
requirement and constitute the most reliable screening constraints at the
moment.

In conclusion, the competition between Eq. $\left( \ref{seffgamow}\right) $
and Eq. $\left( \ref{seffzero}\right) $ is an unequal one. Just a single
term of the former yields a more accurate $EAF$ than a great number of terms
of the latter. Moreover, Eq.$\,\left( \ref{seffgamow}\right) $ takes into
account temperature effects in a consistent and accurate way, whereas Eq. $%
\left( \ref{seffzero}\right) $ practically disregards them. Finally, using
Eq.$\,\left( \ref{seffgamow}\right) $ we were able to modify the reaction
rate formula so that it automatically takes into account the laboratory
electron screening enhancement. The combination of the plasma and laboratory
screening led to the derivation of a combined screening factor which allows
theoreticians to have full control over the theoretical models used for the
description of the electron screening effect.

{\bf ACKNOWLEDGMENTS}

The author is a "Marie Curie" Fellow supported by the European Commission.

\end{document}